\documentclass[aps, prx, 10pt, amssymb,amsmath,superscriptaddress,tightenlines,twocolumn,notitlepage] {revtex4-2}
\usepackage{graphicx}
\usepackage{amsmath}
\usepackage{amssymb}
\usepackage{bm}
\usepackage{color}
\usepackage{hyperref}
\usepackage{tabularx}
\usepackage{multirow}
\usepackage{booktabs}
\usepackage{braket}
\usepackage{epstopdf}

\usepackage[color=orange!60,textsize=scriptsize]{todonotes}
\begin{document}

\title{Transfer learning electronic structure: \\millielectron volt accuracy for sub-million-atom moiré semiconductor}

\author{Ting Bao}\thanks{These authors contributed equally to this work.}
\affiliation{Department of Physics and Astronomy, University of Tennessee, Knoxville, TN 37996, USA}
\affiliation{Department of Physics, Tsinghua University, Beijing 100084, China}
\

\author{Ning Mao}\thanks{These authors contributed equally to this work.}
\affiliation{Max Planck Institute for Chemical Physics of Solids, 01187, Dresden, Germany}


\author{Wenhui Duan}
\affiliation{Department of Physics, Tsinghua University, Beijing 100084, China}

\author{Yong Xu}
\affiliation{Department of Physics, Tsinghua University, Beijing 100084, China}

\author{Adrian Del Maestro}
\affiliation{Department of Physics and Astronomy, University of Tennessee, Knoxville, TN 37996, USA}
\affiliation{Min H. Kao Department of Electrical Engineering and Computer Science, University of Tennessee, Knoxville, Tennessee 37996, USA}

\author{Yang Zhang}
\email{yangzhang@utk.edu}
\affiliation{Department of Physics and Astronomy, University of Tennessee, Knoxville, TN 37996, USA}
\affiliation{Min H. Kao Department of Electrical Engineering and Computer Science, University of Tennessee, Knoxville, Tennessee 37996, USA}


\begin{abstract}
The integration of density functional theory (DFT) with machine learning enables efficient \textit{ab initio} electronic structure calculations for ultra-large systems. In this work, we develop a transfer learning framework tailored for long-wavelength moiré systems. To balance efficiency and accuracy, we adopt a two-step transfer learning strategy: (1) the model is pre-trained on a large dataset of computationally inexpensive non-twisted structures until convergence, and (2) the network is then fine-tuned using a small set of computationally expensive twisted structures. Applying this method to twisted MoTe$_2$, the neural network model generates the resulting Hamiltonian for a 1000-atom system in 200 seconds, achieving a mean absolute error below 0.1 meV. To demonstrate $O(N)$ scalability, we model nanoribbon systems with up to 0.25 million atoms ($\sim9$ million orbitals), accurately capturing edge states consistent with predicted Chern numbers. This approach addresses the challenges of accuracy, efficiency, and scalability, offering a viable alternative to conventional DFT and enabling the exploration of electronic topology in large scale moiré systems towards simulating realistic device architectures. 
\end{abstract}

\maketitle
\section{Introduction}
Twisted bilayer structures are engineered by stacking two-dimensional van der Waals (vdW) layers with a relative rotation, known as the twist angle, producing long-range periodic moiré patterns~\cite{bistritzer2011moire}. At small twist angles, the resulting coulomb repulsion becomes comparable to kinetic energy, giving rise to a rich landscape of strongly correlated states. These include the fractional quantum anomalous Hall (FQAH) effect~\cite{cai2023signatures,zeng2023thermodynamic,xu2023observation,park2023observation},, fractional quantum spin Hall states~\cite{kang2024evidence}, superconducting phases~\cite{cao2018unconventional,cao2018correlated}, Mott and charge-transfer insulators~\cite{PhysRevLett.121.026402,PhysRevB.102.201115,regan2020mott,tang2019wse2,wang2020correlated,ghiotto2021quantum,li2021continuous,https://doi.org/10.48550/arxiv.2202.02055}.The lattice scale of moiré structures is inversely proportional to the twist angle, meaning small-angle twisted structures can contain millions of atoms per unit cell—posing significant challenges for DFT calculations. In addition to twist angles, strain induced by slight layer misalignments can reshape the moiré lattice, creating localized twist-angle defects such as twistons and moiré solitons~\cite{turkel2022orderly,liu2024imaging,thompson2024visualizing}. These defects resemble the granular electronic structures formed by doping and can significantly influence the material's electronic properties, thereby enriching the phase diagrams of moir\'e structures. Traditional DFT methods often struggle with the large scale of small twist-angle structures and the flexibility of strain~\cite{carr2020electronic}, making comprehensive exploration of these systems challenging.

The advent of machine learning offers a promising alternative for studying electronic structures of large-scale systems~\cite{senior2020improved,zhang2018deep,behler2007generalized,mao2024transfer,jia2024moire,zhang2024polarization}. By constructing neural networks that map atomic configurations to electronic Hamiltonians, it becomes practical to efficiently calculate the band topology of large-scale material systems~\cite{li2022deep,gong2023general,zhong2023transferable}. Specifically, while conventional DFT methods scale as $\mathcal{O}(N^3)$ with the number of atoms $N$, machine learning reduces the complexity to linear scaling, $\mathcal{O}(N)$, when sparse diagonalization of the Hamiltonian is performed for bands near the Fermi level.
However, challenges arise when dealing with moir\'e systems, where bandwidths are typically on the order of  1$\sim$10 meV. Training a neural network solely on non-twisted structures risks insufficient accuracy for twisted systems, while exclusive training on twisted structures enhances precision but demands substantial computational resources, reducing efficiency. Balancing these approaches is key to efficiently and accurately exploring moiré systems. Transfer learning provides a powerful solution by leveraging knowledge from non-twisted systems to bypass costly computations~\cite{mao2024transfer}. With minimal inclusion of twisted structures (less than 0.5\%) in the dataset, the fine-tuned neural network can accurately predict the electronic properties of twisted systems. This method mitigates the accuracy-efficiency tradeoff in DFT, paving the way for more extensive investigation of moiré systems.

\begin{figure*}[htbp]
\includegraphics[width=1.8\columnwidth]{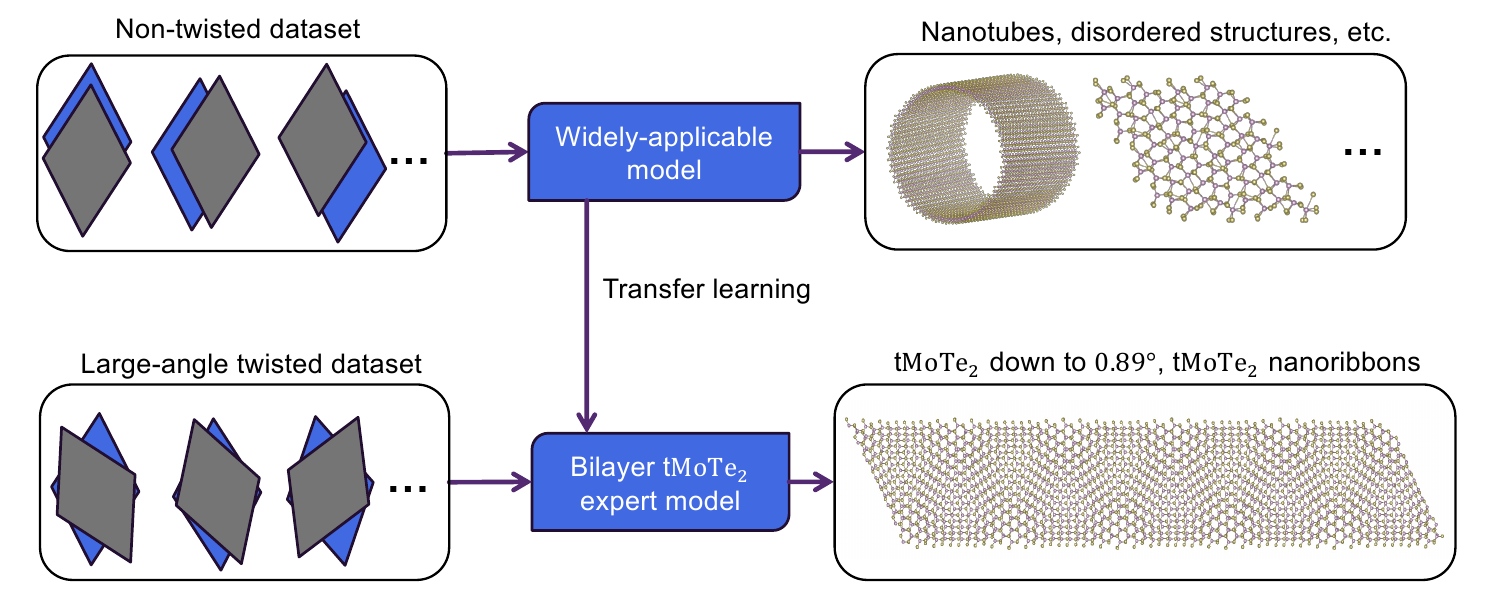}
    \caption{Scheme of transfer learning for electronic structure prediction: A neural network trained solely on a non-twisted dataset, named as widely-applicable model, is capable of predicting the Hamiltonian of large-scale systems, such as nanotubes and disordered structures, but struggles with accurately predicting twisted structures. However, a neural network that undergoes further transfer learning with a training set containing large-angle twisted structures, named as bilayer tMoTe$_2$ expert model, shows a significantly improved ability to predict the Hamiltonian of twisted structures like twsited nanoribbons accurately.
}\label{scheme}
\end{figure*}

In this work, we combine first-principles simulations with machine learning to study the electronic properties of twisted MoTe$_2$. Our simulations span a wide range of twist angles, from $1.89^\circ$ with DFT to $0.89^\circ$ using transfer learning. Initially, we pre-trained a neural network model on a non-twisted dataset, which performed well for large-angle twisted structures but poorly for small-angle twisted structures. Using transfer learning, we incorporated a small amount of large-angle twisted data to fine-tune the model, enabling accurate predictions for small-angle twisted structures. To evaluate the model's ability to predict band topology, we compared the Chern numbers derived from the C$_3$ symmetry eigenvalues. We found an excellent agreement between the results of DFT calculations and the transfer learning neural network, which is consistent with recent experimental findings for twist angles between $5.08^{\circ}$ and $2.13^{\circ}$~\cite{cai2023signatures,zeng2023thermodynamic,park2023observation,xu2023observation,kang2024evidence}. Furthermore, we test the performance of our transfer learning model on twisted structures with varying strains and find that the predictions are in excellent agreement with computationally expensive DFT calculations. Remarkably, the transfer learning model can be extended to ultra-large twisted nanoribbons, containing up to sub-million atoms in a super cell.

\begin{figure*}[htbp]
\includegraphics[width=2\columnwidth]{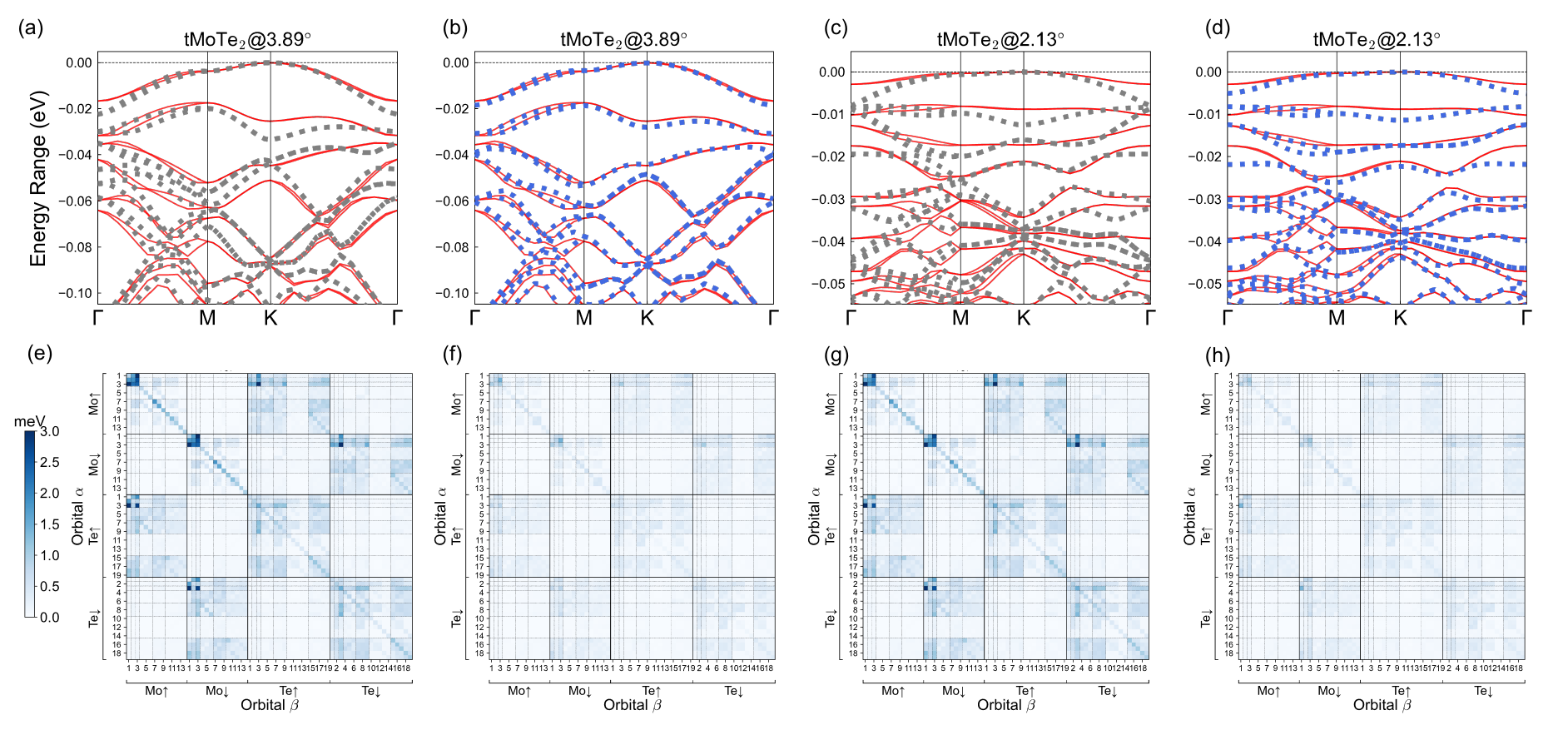}
\caption{
DFT-calculated band structures (red curves) and predicted band structures (gray and blue dots) for twisted MoTe$_2$ with twist angles of (a), (b) 3.89$^\circ$ and (c), (d) 2.13$^\circ$. Panels (a) and (c) show the predictions before transfer learning (gray dots), while panels (b) and (d) show the predictions after transfer learning (blue dots). The mean absolute error (MAE) of $H_{i\alpha, j\beta}$ for different orbitals is presented in panels (e) and (g) for twist angles of 3.89$^\circ$ and 2.13$^\circ$, respectively, before transfer learning, and in panels (f) and (h) after transfer learning.
}\label{compare}
\end{figure*}


\section{Transfer learning process}
We employ pseudo atomic orbitals (PAOs) as the basis for the machine learning training data calculations~\cite{openmx_basis} due to their more compact and efficient representation compared to plance waves. Moreover, PAOs maintain a system-independent gauge, making them particularly suitable for considering the effects of different symmetries. The energy eigen equation involving PAOs is:
\begin{equation}
    H_{i\alpha ,j\beta}(k) \psi_{ n k} = E_n(k) S_{i\alpha ,j\beta}(k) \psi_{nk}
\end{equation}
where the $i/j$ and $\alpha/\beta$ denote the atom and orbital index and the sparse Hamiltonian matrix H$_{i\alpha ,j\beta}$ captures the essential electronic hopping elements within a cutoff radius $R_c$. The overlap matrix S$_{i\alpha ,j\beta}$ reflects the non-orthogonality of the PAO basis and can be constructed without a self-consistent calculation.

The neural network architecture employed in this work is adopted from DeepH-E3~\cite{gong2023general} that can be divided into an input embedding layer, message-passing (MP) layers involving vertex and edge update, and an output E3-linear layer~\cite{weiler20183d}. The input layer begins with the construction of initial vertex features based on the atomic numbers of the atoms, effectively capturing the identity of each atom in the network~\cite{schutt2018schnet}. For the initial edge features, the inter-atomic distances are expanded using a Gaussian basis centered at atomic positions. This approach enables a smooth representation of the distances and the effective propagation of information through the network. Self-loop update, layer normalization and batch normalization are employed to improve the NN performance.

The MP intermediate layers include distinct vertex and edge update blocks that are responsible for iteratively refining the vertex and edge features during the MP process. The vertex update block focuses on updating the features of the vertices based on the incoming messages from neighboring vertices as well as locally (self-loops), while the edge update block propagates edge features by considering both the connected vertices and their corresponding edges~\cite{li2022deep}. For training, we employ the Adam optimizer, initializing the learning rate at 0.002 and iteratively reducing based on a ReduceLROnPlateau schedule when validation model performance plateaus. This dynamic adjustment helps in fine-tuning the model parameters and avoids over-fitting. Self-loops are integrated in the gragh NN to ensure that each vertex can retain its own information during the update process and maintain E3 equivalence to respect the symmetry properties of the data. Crucially, spin-orbit coupling effects are included to accurately capture the electronic properties of materials like tMoTe$_2$. The hyperparameters for these operations are carefully selected to ensure model robustness and accuracy in predicting material properties.

A transfer learning~\cite{pan2009survey} scheme, described in Fig.~\ref{scheme}, allows us to build a powerful and precise model capable of accurately describing twisted systems. Initially, we pre-train our model on a large dataset of non-twisted structures, like steps in the previous DDHT work~\cite{bao2024deep}, allowing it to learn dominant foundational features of electronic materials that are relevant to twisted systems. Since the non-twisted dataset has a simpler structure, both the dataset generation and the training process are significantly faster. Following this, the pre-trained model is fine-tuned on our twisted dataset, enabling it to adapt and specialize to the specific complexities of mori\'e systems. Remarkably, we find that incorporating only 48 twisted structures during the fine-tuning step is sufficient to train a highly effective neural network model. Additionally, by avoiding the need to construct a more complex neural network from the dataset, our transfer learning method mitigates the risk of over-fitting, making the model more generalized to a wide range of twist angles.

\section{Performance Improvement After Transfer Learning}
As a benchmark, we evaluate the performance of the pre-trained neural network model on twist angles of 3.89$^\circ$ and 2.13$^\circ$.

Figs.~\ref{compare}(a) and \ref{compare}(c) show the predicted band structures (dots) and those calculated from DFT methods (lines). While the inherent double degeneracy resulting from the C$_3$ symmetries along the $M-K$ and $K-\Gamma$ paths is preserved, attributed to the advanced equivariant DeepH-E3 neural network framework, a qualitative mismatch is observed. This model discrepancy is quantified via large high Mean Absolute Error (MAE), up to 3 meV for the Mo-Mo block, as seen in Figs.~\ref{compare}(e) and \ref{compare}(g). Moreover, this error margin increases as the twist angles decreases, suggesting a potential systematic issue with the model’s ability to accurately predict Hamiltonian matrix elements for small twist angles, arising from a difficulty to describe long-range effects like the Hartree term.

To improve performance, we perform the transfer learning process described above and observe that this adaptation significantly enhances model accuracy in predicting the Hamiltonian matrix elements for twisted MoTe$_2$ systems. As shown in Figs.~\ref{compare}(f) and \ref{compare}(h), the maximum elementwise MAE is drastically reduced to around 0.5 meV. This suggests that transfer learning improves the model’s ability to generalize across different configurations,better capturing the long-range interactions present in twisted MoTe$_2$ at small angles. Furthermore, the corresponding predicted band structures, presented in Figs.~\ref{compare}(b) and \ref{compare}(d), show agreement with the DFT calculated band structures, underscoring the success of transfer learning in improving model performance.

\section{Prediction of Chern Numbers}
As an application, we exploit the improved accuracy to calculate the Chern number from the NN-inferenced Hamiltonian of tMoTe$_2$. The existence of $C_{3}$ symmetry in twisted MoTe$_2$ allows us to find the Chern number~\cite{wu_topological_2019} at a relatively low cost. At the symmetry-invariant points ($C_3\boldsymbol{k}\rightarrow \boldsymbol{k} + \boldsymbol{K}$), the symmetry eigenvalues of $C_3$ can be defined as:
\begin{equation}
\begin{aligned}
     \xi_{\boldsymbol{k}}=& \left\langle\Psi_{\hat{C_3} \boldsymbol{k}}|\hat{C_3}| \Psi_{\boldsymbol{k}}\right\rangle\\
    =&\sum_{\alpha \beta \boldsymbol{R}} U_{\hat{C_3} \boldsymbol{k}}^{*}(\beta) U_{\boldsymbol{k}}(\alpha) e^{i \left(\hat{C_3} \boldsymbol{k} \tau_{\gamma}-\boldsymbol{k} \cdot \tau_{\alpha} \right)}   \left\langle\boldsymbol{R}, \tau_{\gamma}|\hat{C_3}| \boldsymbol{R}, \tau_{\beta }\right\rangle
\end{aligned}
\end{equation}
where $ \left\langle\boldsymbol{R}, \tau_{\gamma}|\hat{C_3}| \boldsymbol{R}, \tau_{\beta }\right\rangle$ is the $\boldsymbol{k}$-independent representation of the $C_3$ symmetry and $\left | \Psi_{\boldsymbol{k}}\right\rangle= \sum_{\alpha}  U_{\boldsymbol{k}} (\alpha) \left|\boldsymbol{k}, \tau_{\alpha}\right\rangle $is the wavefunction of the crystal. Consequently, at these points (e.g., $\Gamma$, K, K$^{\prime}$), bands are always doubly degenerate, distinguishable by their $C_{3z}$ eigenvalues, $\omega$, satisfying $\omega^3 = (-1)^f$. Here, $f = 0$ and $f = 1$ correspond to spinless and spinful representations, respectively. The Chern number $C$ can be determined from the product of eigenvalues at symmetry-invariant points:
\begin{equation}
e^{i \frac{\pi}{3} C}=\xi_{\Gamma} \xi_{{K}} \xi_{ {K}^{\prime}}
\end{equation}
where $\xi_{ {\Gamma}/\mathrm{K}/{\mathrm{K}^{\prime} }}$ is the spinful $C_3$ eigenvalue at point $\mathrm{\Gamma}/\mathrm{K}/{\mathrm{K}^{\prime}}$.

The supplemental material provides the $C_3$ eigenvalues at high-symmetry points. Within the twist angle range of 5.08$^\circ$ to 1.89$^\circ$, it is observed that for the uppermost three bands, two consistently exhibit eigenvalues of $(\pi/3, \pi, \pi)$ and $(\pi, -\pi/3, -\pi/3)$, both corresponding to a Chern number of +1. The third band, however, shows varying eigenvalues across different twist angles: $(\pi/3, \pi, \pi)$ for angles between 5.08$^\circ$ and 3.15$^\circ$, $(\pi/3, \pi/3, \pi/3)$ at 2.88$^\circ$, and $(-\pi/3, \pi/3, \pi/3)$ for angles between 2.65$^\circ$ and 1.89$^\circ$. These variations indicate that the third band undergoes two distinct topological phase transitions: the first occurs at the $K$ point at 2.88$^\circ$, followed by a second transition at the $\Gamma$ point at 2.65$^\circ$.

\begin{figure}[htbp]
\includegraphics[width=1.0\columnwidth]{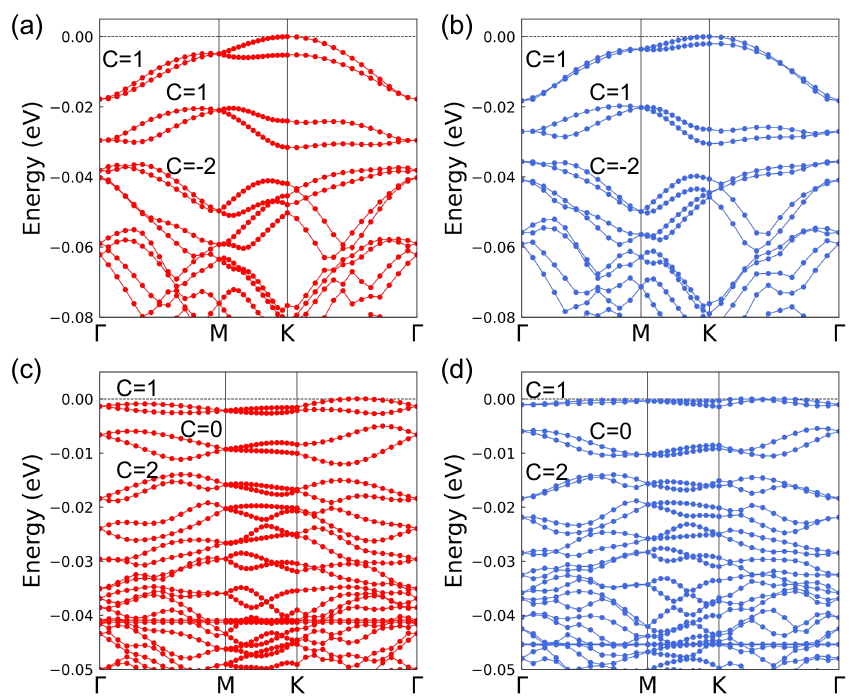}
\caption{
(a), (b) Band structures for a twist angle of 3.68$^\circ$ with a strain of 0.006, where (a) is calculated using OpenMX, and (b) is predicted by the transfer learning neural network, both results having Chern number 1,1,-2 for the first three bands respectively.
(c), (d) Band structures for a twist angle of 2.05$^\circ$ with a strain of 0.01, where (c) calculated using OpenMX, and (d) is predicted by the transfer learning neural network, both results having Chern number 1,0,2 for the first three bands respectively.
}
\label{strain}
\end{figure}

\section{Extension to strained moir\'e systems}
In experimental studies of mori\'e systems, subtle strain effects arising during the fabrication process can significantly impact the electronic properties of materials ~\cite{kerelsky2019maximized,artaud2016universal,huder2018electronic}. 
For bi-layers, when both lattices are equally affected, the strain is termed homostrain, whereas differential strain between the two lattices is known as heterostrain. Homostrain of a certain magnitude causes a corresponding shift in the moir\'e wavelength along the direction of the strain, but does not lead to significant variations in the moir\'e wavelength. Given the typical sub-percent strains observed in experiments~\cite{liu2024imaging,thompson2024visualizing}, the effect of homostrain on moir\'e wavelengths can be neglected. Heterostrain, on the other hand, has a substantial impact on moir\'e wavelengths. 

Conventional DFT calculations often struggle to accurately handle strained twisted systems due to the complexity and uncertainty in the magnitude and direction of the strain, making these calculations time-consuming and potentially inaccurate. Our model, in addition to being applicable to standard twisted systems, can also be extended to strained twisted systems, offering a robust alternative to traditional methods. To validate its effectiveness, we compared its predictions for stained tMoTe$_2$ at 3.68$^\circ$ and 2.05$^\circ$ as shown in Fig.~\ref{strain}. The predicted band structure of the strained system closely matches calculations obtained from DFT package OpenMX, demonstrating the robustness of our NN model. Notably, the influence of strain extends beyond structural modifications, inducing significant changes in the topological properties of twisted systems. This effect is particularly evident at smaller twist angles, where lower energy scales dominate. As shown in Fig.~\ref{strain}(a)(b), the 3.68$^\circ$-twisted structure with a strain of 0.006 exhibits Chern numbers of 1, 1, and -2 for its first three bands, consistent with those of the unstrained 3.89$^\circ$- and 3.48$^\circ$-twisted cases. In contrast, the 2.05$^\circ$-twisted structure under similar strain displays Chern numbers of 1, 0, and 2 for its first three energy bands, differing from the unstrained 2.28$^\circ$ structure, which has Chern numbers of 1, 1, and -2. This divergence arises because smaller twist angles correspond to lower energy scales, making topological properties more sensitive to strain and global atomic configuration changes.

This result underscores the strain-induced topological transitions present in such systems and is well captured by our NN model.


\section{Scalability Performance}

\begin{figure*}[htbp]
\includegraphics[width=2.0\columnwidth]{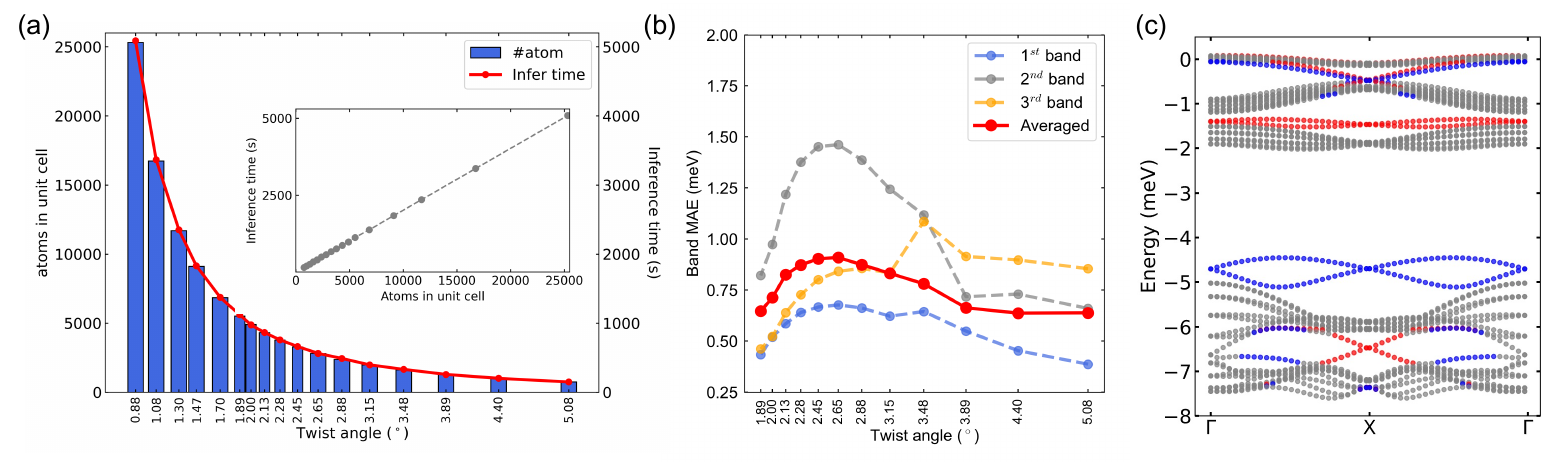}
\caption{
(a) The inference time for the transfer learning neural network model, with the blue bar plot representing the number of atoms at twist angle from 5.08$\circ$ to 0.88$\circ$ and the red line indicating the inference time. The insertion shows the linear relation between inference time and number of atoms.
(b) The mean absolute error (MAE) of the top three valence bands along k-path $\Gamma$-M-K-$\Gamma$ for different twist angles, ranging from 5.08$^\circ$ to 1.89$^\circ$.  (c) The band structure of nanoribbon at twist angle of 0.88$^\circ$, where the nanoribbon is periodic along the lattice vector b and finite along the the lattice vector a, spanning a length of 5 unit cells. The edge states of spin-up/-down are in red/blue, and the bulk states are in gray.}
\label{infer1}
\end{figure*}

Unlike traditional DFT calculation with $O(N^3)$ complexity, our transfer learning NN workflow maintains $O(N)$ complexity at both the Hamiltonian inference step and Hamiltonian matrix diagonalization (due to the resulting sparse nature), allowing for application to ultra-large systems. Fig.~\ref{infer1}(a) shows the NN inference time for tMoTe$_2$ at different twist angles,  with the $O(N)$ complexity illustrated in the inset by the linear relation between the inference time and the number of atoms. Importantly, the prediction accuracy persists with the increasing system size as shown in Fig.~\ref{infer1}(b), where the averaged MAE error along a $k$ path of the first 3 bands is always at sub-meV to 1 meV level from 5.08$^\circ$ to 1.89$^\circ$, as validated by DFT benchmarks. The NN results also accurately capture the trend of decreasing bandwidth in the first band with decreasing twist angle, as depicted in Fig.~\ref{infer1}(b) explaining the reduction of MAE as small angles. To diagonalize the predicted Hamiltonian matrices in \textit{k}-space, we first select an initial estimate for the energy horizon, which can be determined from small twist-angle systems. Subsequently, we focus on extracting a specific number of eigenvalues around this energy horizon. By employing a partial diagonalization approach for the sparse matrix, facilitated by the Pardiso package, we achieve the computation of energy eigenvalues with an $O(N)$ complexity. Systematic tests have been conducted across various system scales, with the corresponding memory and runtime consumption presented in Table.~\ref{FigS3}-\ref{FigS6}.

Putting everything together, the resulting high-efficiency workflow enables calculations on sub-million-atom systems such as a nanoribbon system that includes spin-orbit coupling (SOC) effects. We inference Hamiltonians of the 0.88$^\circ$-twisted nanoribbon as examples, the largest nanoribbon system here is 10 times of a single 0.88$^\circ$-twisted unit cell, comprising 253,140 atoms (8,775,520 orbitals). And we also conduct edge states projection on a 5-time-unitcell 0.88$^\circ$-twisted nanoribbon as it band structure shown in Fig.~\ref{infer1}(c), where the edge states of spin-up/-down are in red/blue and the bulk states are in gray. The edge states align with the Chern number of the periodic 0.88$^\circ$-twisted structures. Specifically, gapless edge states are observed between the original first band (Chern number -1) and the second band (Chern number 1), whereas gapped edge states are present between the original second band (Chern number 1) and the third band (Chern number 0).


Furthermore, for TMD moiré systems where SOC is negligible, such as those dominated by Gamma-valley bands at the Fermi level, we can scale up to systems of 0.5 million atoms (twice the size of the SOC case). For simpler twisted graphene or hBN systems, our approach enables scaling to millions of atoms, surpassing the upper limits of traditional DFT methods.

\section{Conclusion}
In summary, we have proposed a transfer learning framework capable of handling small-angle twisted mori\'e structures with an inference complexity that scales linearly with the number of atoms in the unit cell. The utility and accuracy of the method is demonstrated by comparing to a neural network model exclusively on a non-twisted dataset. By fine-tuning with a small large-angle twisted dataset, we achieve both high efficiency and sub-meV accuracy for model predictions of electronic Hamiltonians. To further validate the utility of the predicted Hamiltonian, we calculate band Chern numbers based on the C$_3$ symmetry eigenvalues and find that the predicted band topology is consistent with experimental results. Our training dataset only included structures with C$_3$ symmetry, however we find that our model also accurately predicts strained moir\'e systems that explicitly break these symmetries. Remarkably, our model also demonstrates excellent scalability, making it capable of predicting ultra-large systems approaching real devices, such as twisted nanoribbons with sub-million atoms.
The transfer learning predicted Hamiltonian can not only be used to calculate the band topology of the topmost few bands but also to explore strong correlations at different fillings, paving the way for further many-body calculations~\cite{xu2024multiple}.

\section*{Acknowledgments}
We thanks Jingheng Fu, He Li, Zilong Yuan, Mark Gates, Natalie Beams and Ahmad Abdelfattah for helpful discussions on neural network and large matrix diagonalization, Shiyong Wang and Tingxin Li for experimental collaboration on stained MoTe$_2$. N.M. acknowledges the financial support from the Alexander von Humboldt Foundation. Y.Z. is supported by the start up fund at University of Tennessee.

\bibliography{ref}
\clearpage
\pagebreak
\onecolumngrid
\begin{center}
\textbf{\large Supplemental Materials}
\end{center}
\setcounter{figure}{0}
\renewcommand{\thefigure}{S\arabic{figure}}
\setcounter{equation}{0}
\renewcommand{\theequation}{S\arabic{equation}}
\setcounter{table}{0}
\renewcommand{\thetable}{S\arabic{table}}

\subsection{Details of transfer learning lattice relaxation}

To achieve lattice relaxation, we employ a two-step transfer learning approach, using the DeePMD-kit code~\cite{zhang2018deep}. Initially, we construct $3 \times 3 \times 1$ MM, MX, and XM configurations, along with 28 distinct intermediate transition states. The training dataset for the initial lattice relaxation is composed of 200 random perturbations applied to these transition states, resulting in a total of 6200 different structures. We utilize the collected energy, force, and virial data to train the lattice relaxation model, leveraging the DeepPot-SE descriptor. In the second step, the model is augmented with a small set of twisted structures, and transfer learning techniques are applied. By freezing the parameters of the embedding layers and concentrating on the hidden and output layers, we develop a transfer learning neural network capable of relaxing twisted MoTe$_2$ with twist angles ranging between $5.89^{\circ}$ and $1.89^{\circ}$.

\subsection{Details of OpenMX Calculations}

In our OpenMX calculations, we use Projected Atomic Orbitals (PAOs) with the basis sets specified as \textit{Mo7.0-s3p2d1} and \textit{Te7.0-s3p2d2}~\cite{openmx_basis,openmx_largescale}. The notation \textit{7.0} indicates a cutoff radius of 7.0 Bohr. For the \textit{Mo7.0-s3p2d1} basis set, \textit{s3p2d1} corresponds to 3 sets of $s$-orbitals, 2 sets of $p$-orbitals, and 1 set of $d$-orbitals, comprising a total of 14 atomic orbitals. Similarly, \textit{Te7.0-s3p2d2} includes 3 sets of $s$-orbitals, 2 sets of $p$-orbitals, and 2 sets of $d$-orbitals, resulting in 19 atomic orbitals.
To obtain the overlap matrices for twist angles between $5.08^\circ$ and $0.89^\circ$, we employed the PBE exchange-correlation functional and norm-conserving pseudopotentials~\cite{openmx_pseudopotential}, the overlap matrices are calculated at a negligible low cost.

\subsection{Details of transfer learning electronic Hamiltonian}

Both the first-step and two-step neural network models are trained based on the DeepH-E3 architecture. The initial training dataset comprises 576 non-twisted MoTe$_2$ supercells of size $6 \times 6 \times 1$, while the twisted dataset includes 48 twisted MoTe$_2$ configurations with twist angles ranging between $9^\circ$ and $21^\circ$. The representation for the initial vertex (edge) features is $64\times0e$, while for the intermediate features, it is $64\times0e + 32\times1o + 32\times2e + 16\times3o + 8\times4e + 4\times5o$. Spherical harmonics with $l = 0$ to $5$ are utilized for the material representations. During model training, we utilize a ReduceLrOnPlateau learning scheme, starting with an initial learning rate of 0.002. Specifically, the learning rate is halved if no improvement is observed for 100 epochs, with a cooldown period of 40 epochs before further reductions are considered. The threshold for determining a plateau is set at 0.05, and the process is verbose to provide updates during training. Additionally, the training process includes a safeguard against prolonged increases in validation loss. If the validation loss exceeds twice the best loss for more than 20 epochs, the model reverts to the best checkpoint, and the learning rate is reduced by $ 20\% $, ensuring the model can recover and continue improving. The dataset is split into training, validation, and testing sets, comprising 80\%, 15\%, and 5\% of the total dataset, respectively, with the
batch size of 1. The training process spans at most 5,000 epochs and is terminated early if the validation loss reaches zero within the maximum number of epochs.


\begin{table}[]
\begin{tabular}{@{}cllll@{}}
\toprule
\multicolumn{1}{l}{\textbf{Twist angle}} & \multicolumn{1}{c}{\textbf{Mo-Mo}} & \multicolumn{1}{c}{\textbf{Mo-Te}} & \multicolumn{1}{c}{\textbf{Te-Mo}} & \multicolumn{1}{c}{\textbf{Te-Te}} \\ \midrule
\textbf{5.08}                            & 0.15/0.07(-57\%)                   & 0.23/0.11(-53\%)                   & 0.21/0.10(-52\%)                   & 0.22/0.12(-46\%)                   \\
\textbf{4.41}                            & 0.16/0.07(-57\%)                   & 0.24/0.11(-51\%)                   & 0.21/0.11(-51\%)                   & 0.22/0.12(-45\%)                   \\
\textbf{3.89}                            & 0.16/0.07(-56\%)                   & 0.24/0.12(-51\%)                   & 0.22/0.11(-51\%)                   & 0.22/0.12(-45\%)                   \\
\textbf{3.48}                            & 0.16/0.07(-56\%)                   & 0.25/0.12(-50\%)                   & 0.22/0.11(-50\%)                   & 0.22/0.13(-43\%)                   \\
\textbf{3.15}                            & 0.15/0.07(-56\%)                   & 0.24/0.12(-49\%)                   & 0.21/0.11(-48\%)                   & 0.22/0.13(-41\%)                   \\
\textbf{2.88}                            & 0.15/0.07(-56\%)                   & 0.24/0.13(-48\%)                   & 0.22/0.11(-47\%)                   & 0.22/0.13(-40\%)                   \\
\textbf{2.65}                            & 0.15/0.07(-55\%)                   & 0.24/0.13(-48\%)                   & 0.22/0.12(-47\%)                   & 0.22/0.13(-39\%)                   \\
\textbf{2.45}                            & 0.15/0.07(-55\%)                   & 0.24/0.13(-47\%)                   & 0.22/0.12(-47\%)                   & 0.22/0.13(-39\%)                   \\
\textbf{2.28}                            & 0.15/0.07(-55\%)                   & 0.24/0.13(-46\%)                   & 0.22/0.12(-47\%)                   & 0.22/0.13(-38\%)                   \\
\textbf{2.13}                            & 0.14/0.07(-55\%)                   & 0.24/0.13(-46\%)                   & 0.22/0.12(-47\%)                   & 0.22/0.13(-38\%)                   \\
\textbf{2.00}                            & 0.22/0.14(-34\%)                   & 0.31/0.18(-43\%)                   & 0.25/0.16(-35\%)                   & 0.29/0.24(-19\%)                   \\
\textbf{1.89}                            & 0.14/0.07(-54\%)                   & 0.24/0.13(-45\%)                   & 0.22/0.12(-46\%)                   & 0.22/0.14(-36\%)                   \\ \bottomrule
\end{tabular}
\caption{
The average error of individual elementwise Hamiltonian blocks for twist angles ranging from 5.08$^\circ$ to 1.89$^\circ$. The data are in meV unit and shown as \textit{error before transfer learning} / \textit{error after transfer learning} (\textit{reduced by percentage})
}
\end{table}

\begin{table}[]
\begin{tabular}{@{}cllll@{}}
\toprule
\multicolumn{1}{l}{\textbf{Twist  angle}} & \multicolumn{1}{c}{\textbf{Mo-Mo}} & \multicolumn{1}{c}{\textbf{Mo-Te}} & \multicolumn{1}{c}{\textbf{Te-Mo}} & \multicolumn{1}{c}{\textbf{Te-Te}} \\ \midrule
\textbf{5.08}                             & 4.20/1.57(-62\%)                   & 3.51/1.09(-69\%)                   & 3.78/1.07(-72\%)                   & 1.39/0.68(-51\%)                   \\
\textbf{4.41}                             & 4.02/1.52(-62\%)                   & 3.41/0.97(-72\%)                   & 3.43/1.18(-66\%)                   & 1.42/0.74(-48\%)                   \\
\textbf{3.89}                             & 4.05/1.54(-62\%)                   & 3.38/1.02(-70\%)                   & 3.53/1.06(-70\%)                   & 1.48/0.82(-44\%)                   \\
\textbf{3.48}                             & 3.92/1.51(-62\%)                   & 3.35/1.01(-70\%)                   & 3.37/0.97(-71\%)                   & 1.46/0.82(-44\%)                   \\
\textbf{3.15}                             & 3.69/1.39(-62\%)                   & 3.14/1.23(-61\%)                   & 3.27/1.13(-66\%)                   & 1.50/0.86(-43\%)                   \\
\textbf{2.88}                             & 3.50/1.31(-63\%)                   & 3.08/1.33(-57\%)                   & 3.07/0.91(-70\%)                   & 1.50/0.85(-43\%)                   \\
\textbf{2.65}                             & 3.62/1.32(-64\%)                   & 3.08/1.37(-56\%)                   & 3.27/1.06(-68\%)                   & 1.51/0.91(-40\%)                   \\
\textbf{2.45}                             & 3.47/1.23(-65\%)                   & 3.05/1.46(-52\%)                   & 3.04/0.85(-72\%)                   & 1.48/0.86(-42\%)                   \\
\textbf{2.28}                             & 3.45/1.19(-65\%)                   & 3.02/1.54(-49\%)                   & 3.08/0.88(-72\%)                   & 1.48/0.86(-41\%)                   \\
\textbf{2.13}                             & 3.42/1.14(-67\%)                   & 2.99/1.64(-45\%)                   & 3.01/0.85(-72\%)                   & 1.47/0.86(-42\%)                   \\
\textbf{2.00}                             & 3.76/2.73(-27\%)                   & 2.50/2.13(-15\%)                   & 2.63/1.35(-48\%)                   & 1.96/1.67(-14\%)                   \\
\textbf{1.89}                             & 3.38/1.23(-64\%)                   & 2.93/1.84(-37\%)                   & 3.02/0.94(-69\%)                   & 1.47/0.90(-39\%)                   \\ \bottomrule
\end{tabular}
\caption{
The maximum error of individual elementwise Hamiltonian blocks for twist angles ranging from 5.08$^\circ$ to 1.89$^\circ$. The data are in meV unit and shown as \textit{error before transfer learning} / \textit{error after transfer learning} (\textit{reduced by percentage})
}
\end{table}

\begin{table}[htbp]
\centering
\begin{tabular}{l|c|c|c|c|c|c|c|c|c|c}
\hline
Dense Matrix & 26k & 35k & 45k & 56k & 69k & 83k & 98k& 114k & 131k & 170k\\ \hline
Time (s)    & 31    & 61 & 84 & 130 & 202 &361 &549 &620 & 892&942\\  \hline
Atom number & 762    & 1014 &1302  & 1626 &1986  & 2382 &2814  &3282  & 3786 &4902   \\  \hline
Memory (GB) & 11*2 & 20*2 & 33*2 & 51*2 & 76*2  &109*2 &152*2 &207*2 & 276*2 &462*2\\ \hline
Twist angle ($^o$) & 5.08 & 4.4 & 3.89 & 3.48 & 3.15&2.88&2.65&2.45&2.28 &2.00\\
\hline
\end{tabular}
\caption{Generalized eigenvalue problem tests for bulk structure}
\label{FigS3}
\end{table}

\begin{table}[htbp]
\centering
\begin{tabular}{l|c|c|c|c|c|c}
\hline
Dense Matrix & 26k & 53k & 79k & 106k & 132k & 158k \\ \hline
Time (s)    & 15    & 111 & 264 & 521 & 869 & 886 \\  \hline
Atom number & 762    & 1524 &2286  & 3048 &3810  & 4572   \\  \hline
Memory (GB) & 22 & 89 & 200 & 357 & 558  & 803 \\ \hline
Length  & 1 & 2 & 3 & 4 & 5 & 6\\
\hline
\end{tabular}
\caption{Generalized eigenvalue problem tests for ribbon of twist angle 5.08$^o$}
\end{table}

\begin{table}[htbp]
\centering
\begin{tabular}{l|c|c|c|c}
\hline
Delete element ($\%$) & 0.0 & 0.1 & 0.01 & 0.001 \\ \hline
Time (s)    & 880    & 877 & 878 &  875\\  \hline
Memory (GB) & 803  & 803 & 803 & 803  \\ \hline
\end{tabular}
\caption{Cutoff test for 1*6 ribbon of twist angle 5.08$^o$}
\end{table}

\begin{table}[h!]
\centering
\begin{tabular}{c|c|c|c|c|c}
\hline
\textbf{Twist Angle} & \textbf{ Atoms} & \textbf{Matrix Dimensions} & \textbf{Memory} & \textbf{Computational Time} & \textbf{\# Eigenvalues} \\ \hline
2.88 & 2382  & 82576   & 42GB   & 20min    & 40  \\ \hline
2.65 & 2814  & 97552   & 52GB   & 91min    & 300 \\ \hline
2.00 & 4902  & 169936  & 104GB  & 51min    & 30  \\ \hline
1.89 & 5514  & 191152  & 108GB  & 250min   & 300 \\ \hline
0.88 5*nanoribbon & 126570 & 4387760  & 1518GB  & 960 min   & 50 \\ \hline
0.88 10*nanoribbon & 253140 & 8775520 & 2142 GB & 764 min & 50 \\ \hline
\end{tabular}
\caption{Table summarizing computational parameters for different twist angles and configurations.}
\label{FigS6}
\end{table}

\begin{figure}[htbp]
\includegraphics[width=0.8\columnwidth]{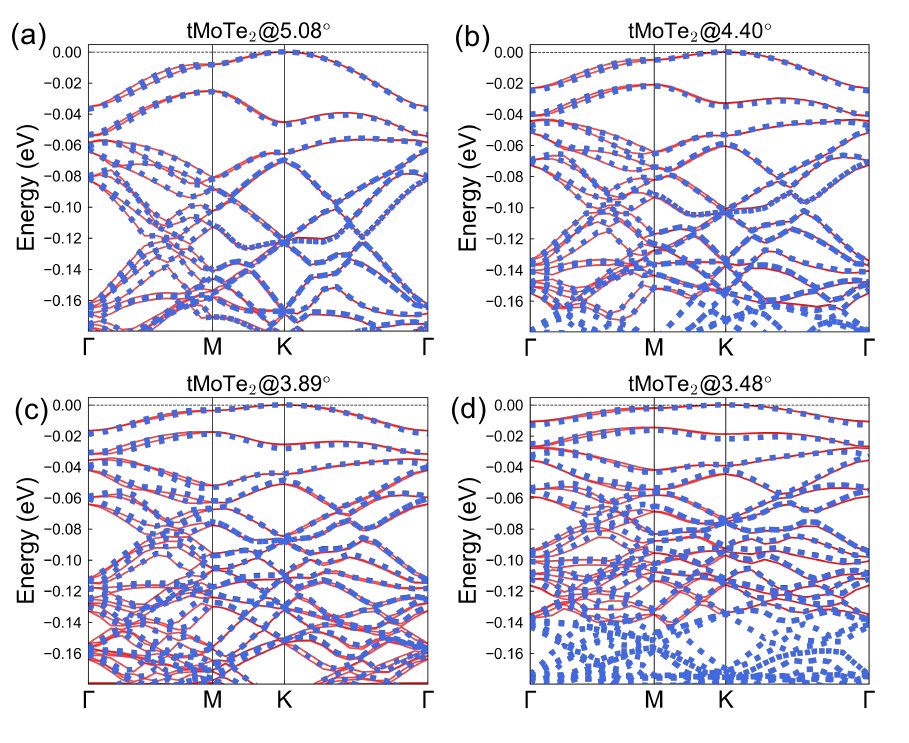}
\caption{
DFT-calculated band structures (red curves) and predicted band structures (blue dots) for twisted MoTe$_2$ with twist angles of (a) 5.08$^\circ$, (b) 4.40$^\circ$, (c) 3.89$^\circ$, and (d) 3.48$^\circ$, respectively.
}
\end{figure}

\begin{figure}[htbp]
\includegraphics[width=0.8\columnwidth]{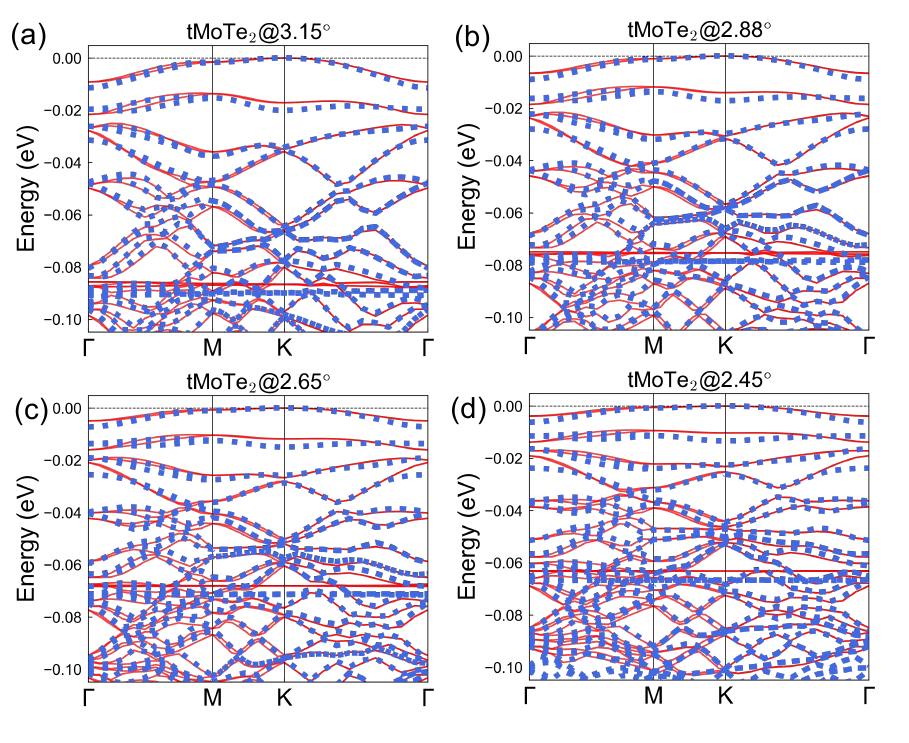}
\caption{
DFT-calculated band structures (red curves) and predicted band structures (blue dots) for twisted MoTe$_2$ with twist angles of (a) 3.15$^\circ$, (b) 2.88$^\circ$, (c) 2.65$^\circ$, and (d) 2.45$^\circ$, respectively.
}
\end{figure}

\begin{figure}[htbp]
\includegraphics[width=0.8\columnwidth]{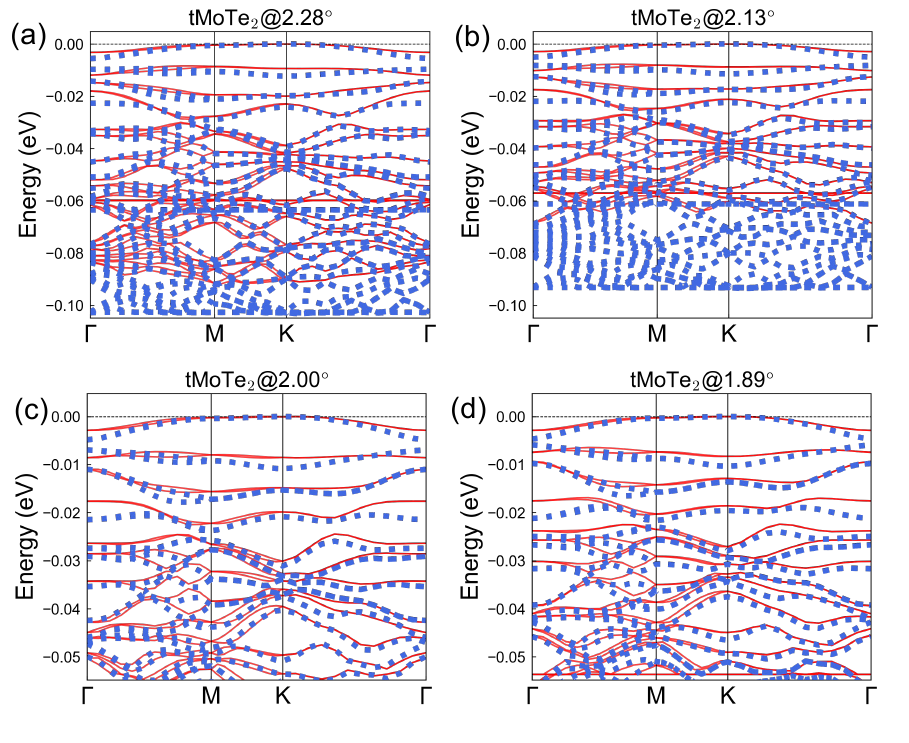}
\caption{
DFT-calculated band structures (red curves) and predicted band structures (blue dots) for twisted MoTe$_2$ with twist angles of (a) 2.28$^\circ$, (b) 2.13$^\circ$, (c) 2.00$^\circ$, and (d) 1.89$^\circ$, respectively.
}
\end{figure}

\begin{figure}[htbp]
\includegraphics[width=0.8\columnwidth]{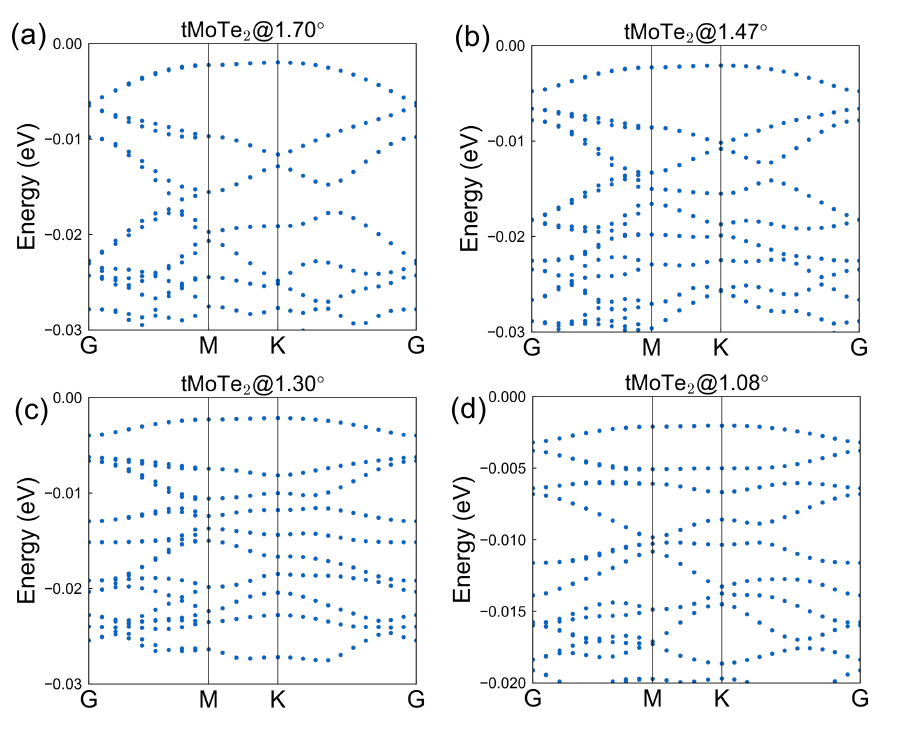}
\caption{
Predicted band structures (blue dots) for twisted MoTe$_2$ with twist angles of (a) 1.70$^\circ$, (b) 1.47$^\circ$, (c) 1.30$^\circ$, and (d) 1.08$^\circ$, respectively.
}
\end{figure}

\begin{figure}[htbp]
\includegraphics[width=0.8\columnwidth]{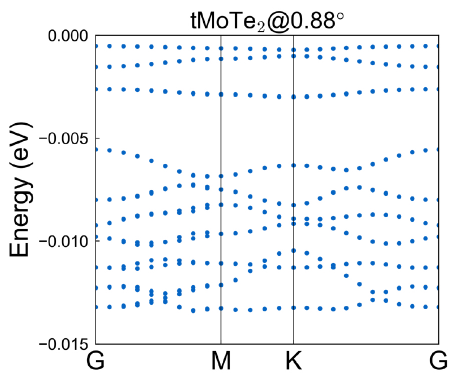}
\caption{
Predicted band structures (blue dots) for twisted MoTe$_2$ with twist angles of 0.88$^\circ$.
}
\end{figure}

\begin{figure}[htbp]
\includegraphics[width=0.8\columnwidth]{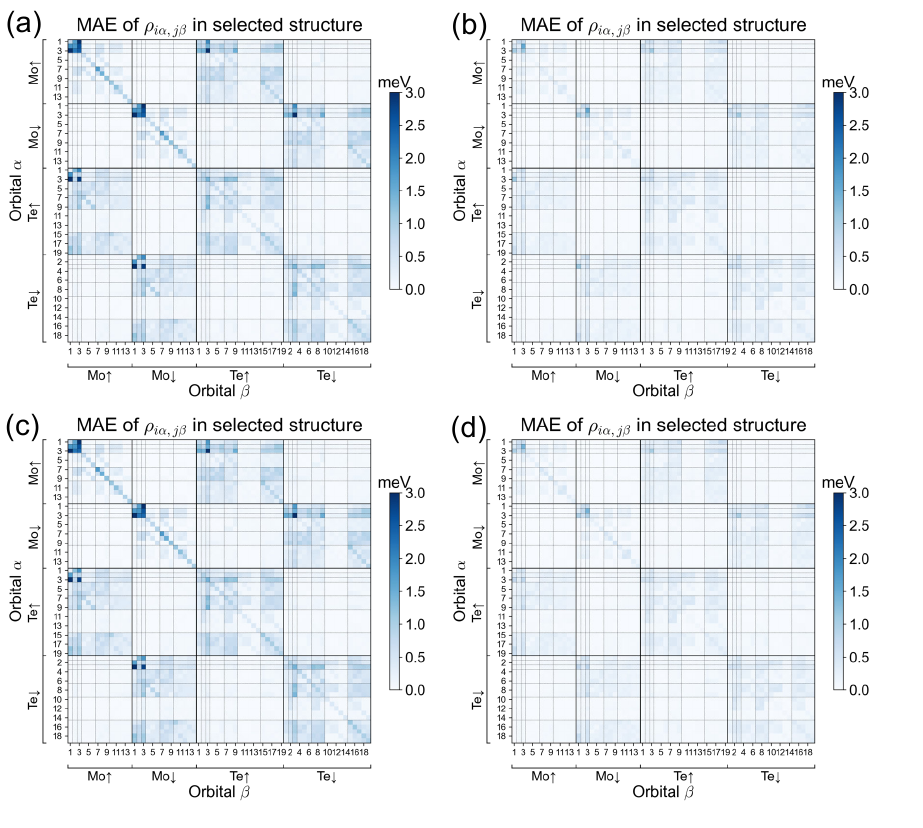}
\caption{
The mean absolute error of $H{i\alpha, j\beta}$ for twist angles of (a), (b) 5.08$^\circ$, and (c), (d) 4.40$^\circ$. Panels (a) and (c) show the predictions before transfer learning, while panels (b) and (d) show the predictions after transfer learning.
}
\end{figure}

\begin{figure}[htbp]
\includegraphics[width=0.8\columnwidth]{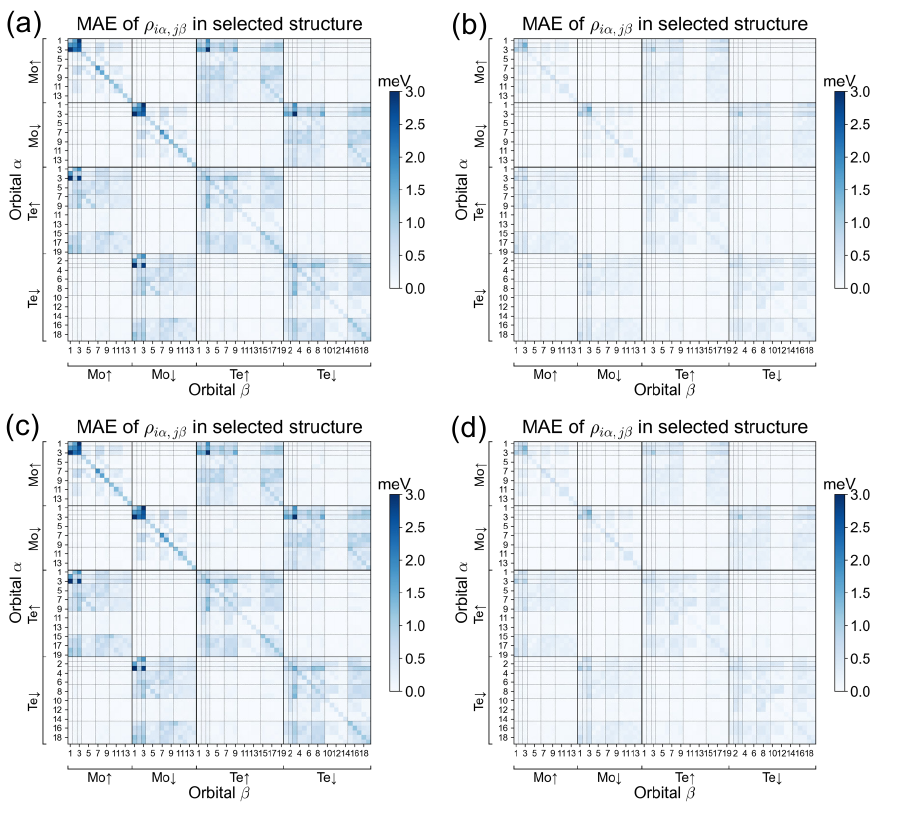}
\caption{
The mean absolute error of $H{i\alpha, j\beta}$ for twist angles of (a), (b) 3.89$^\circ$, and (c), (d) 3.48$^\circ$. Panels (a) and (c) show the predictions before transfer learning, while panels (b) and (d) show the predictions after transfer learning.
}
\end{figure}

\begin{figure}[htbp]
\includegraphics[width=0.8\columnwidth]{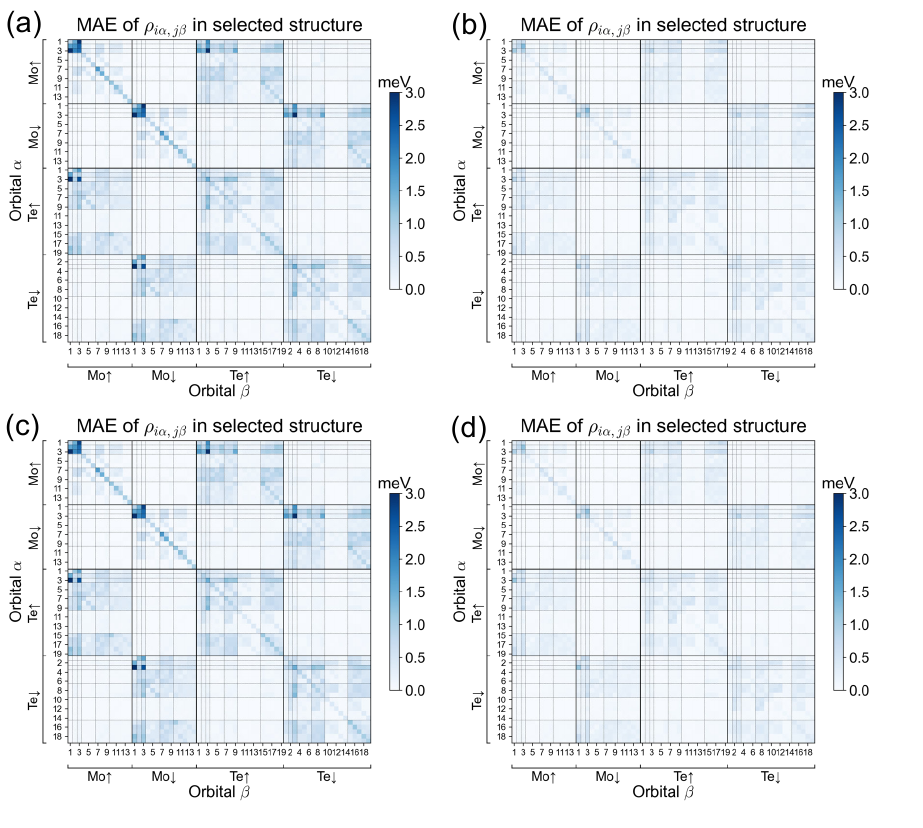}
\caption{
The mean absolute error of $H{i\alpha, j\beta}$ for twist angles of (a), (b) 3.15$^\circ$, and (c), (d) 2.88$^\circ$. Panels (a) and (c) show the predictions before transfer learning, while panels (b) and (d) show the predictions after transfer learning.
}
\end{figure}

\begin{figure}[htbp]
\includegraphics[width=0.8\columnwidth]{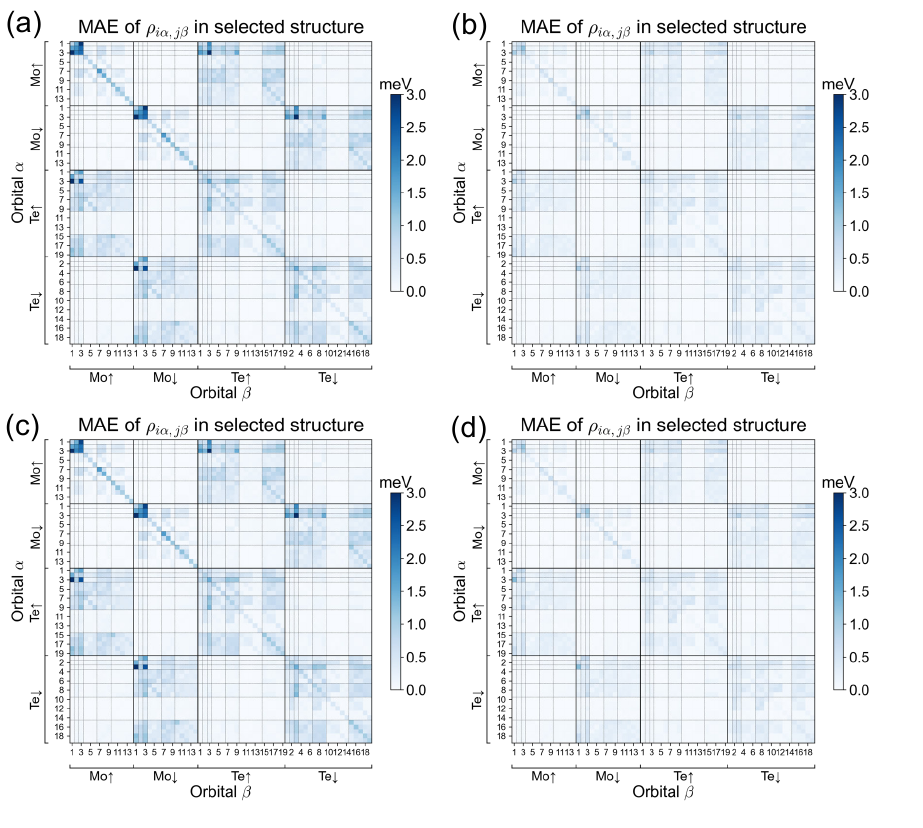}
\caption{
The mean absolute error of $H{i\alpha, j\beta}$ for twist angles of (a), (b) 2.65$^\circ$, and (c), (d) 2.45$^\circ$. Panels (a) and (c) show the predictions before transfer learning, while panels (b) and (d) show the predictions after transfer learning.
}
\end{figure}

\begin{figure}[htbp]
\includegraphics[width=0.8\columnwidth]{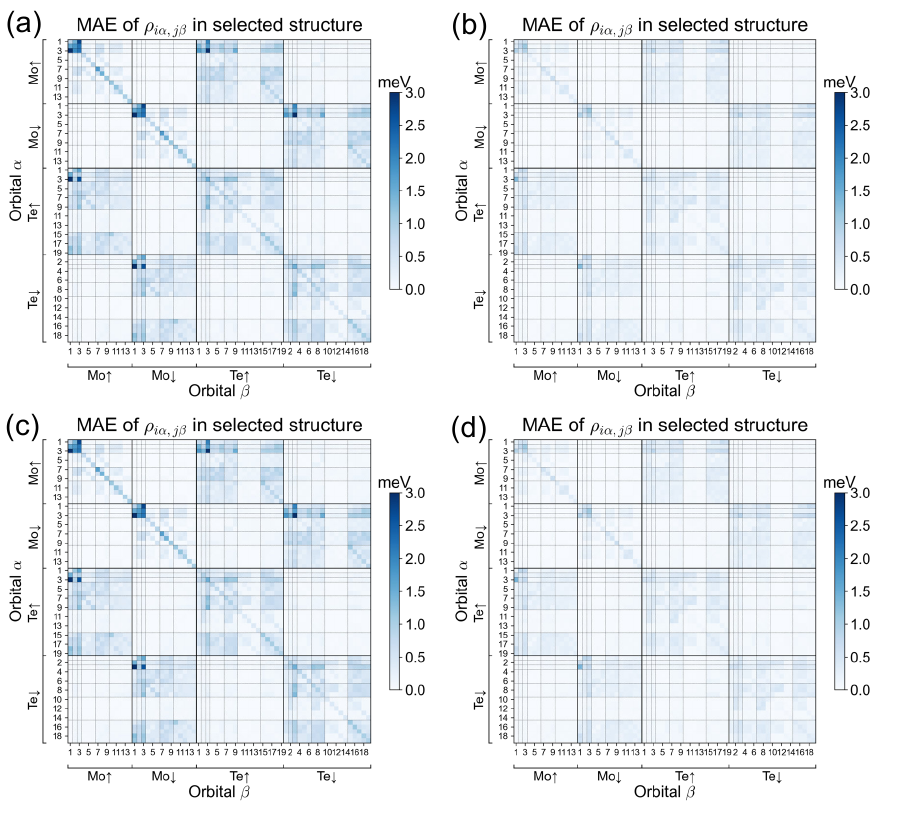}
\caption{
The mean absolute error of $H{i\alpha, j\beta}$ for twist angles of (a), (b) 2.28$^\circ$, and (c), (d) 2.13$^\circ$. Panels (a) and (c) show the predictions before transfer learning, while panels (b) and (d) show the predictions after transfer learning.
}
\end{figure}

\begin{figure}[htbp]
\includegraphics[width=0.8\columnwidth]{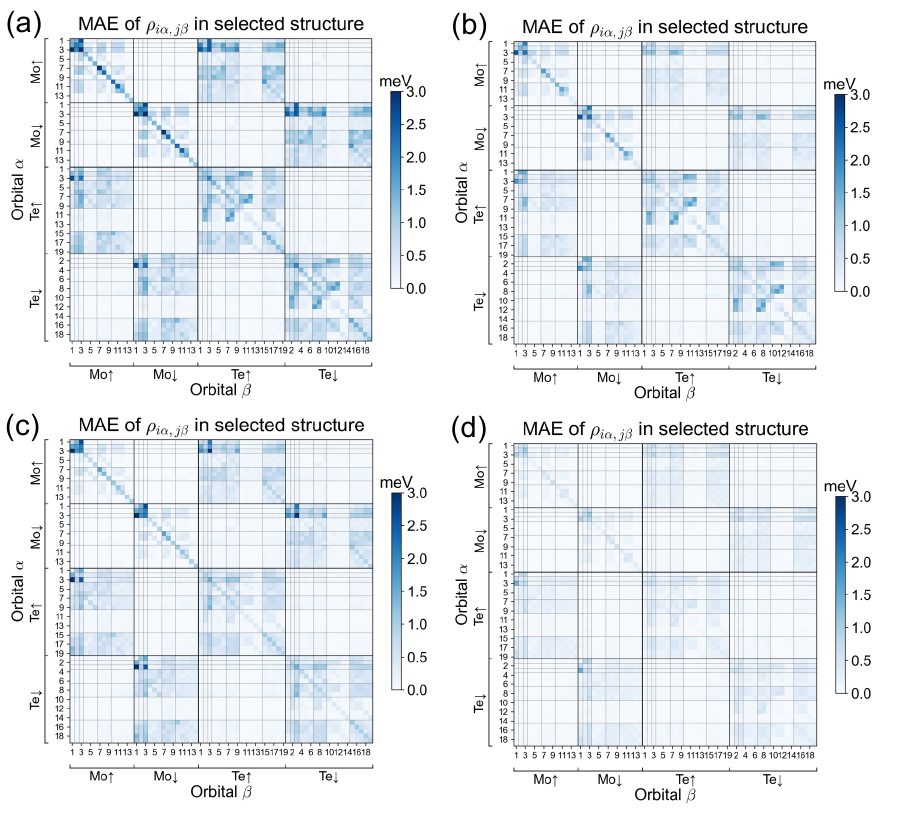}
\caption{
The mean absolute error of $H{i\alpha, j\beta}$ for twist angles of (a), (b) 2.00$^\circ$, and (c), (d) 1.89$^\circ$. Panels (a) and (c) show the predictions before transfer learning, while panels (b) and (d) show the predictions after transfer learning.
}
\end{figure}

\begin{table}[]
\begin{tabular}{@{}ccccc|c|ccccc@{}}
\toprule
\textbf{5.08} & \textbf{$\Gamma$} & \textbf{K} & \textbf{K'} & \textbf{Chern Number (↑)} & \textbf{} & \textbf{2.13} & \textbf{$\Gamma$} & \textbf{K} & \textbf{K'} & \textbf{Chern Number (↑)} \\ \midrule
band 1        & $\pi$/3           & $\pi$      & $\pi$       & 1                         &           & band 1        & $\pi$/3           & $\pi$      & $\pi$       & 1                         \\ \midrule
band 2        & $\pi$             & -$\pi$/3   & -$\pi$/3    & 1                         &           & band 2        & $\pi$             & -$\pi$/3   & -$\pi$/3    & 1                         \\ \midrule
band 3        & $\pi$/3           & $\pi$      & $\pi$       & -2                        &           & band 3        & -$\pi$/3          & $\pi$/3    & $\pi$/3     & 1                         \\ \midrule
\textbf{4.41} & \textbf{$\Gamma$} & \textbf{K} & \textbf{K'} & \textbf{Chern Number (↑)} & \textbf{} & \textbf{2.01} & \textbf{$\Gamma$} & \textbf{K} & \textbf{K'} & \textbf{Chern Number (↑)} \\ \midrule
band 1        & $\pi$/3           & $\pi$      & $\pi$       & 1                         &           & band 1        & $\pi$/3           & $\pi$      & $\pi$       & 1                         \\ \midrule
band 2        & $\pi$             & -$\pi$/3   & -$\pi$/3    & 1                         &           & band 2        & $\pi$             & -$\pi$/3   & -$\pi$/3    & 1                         \\ \midrule
band 3        & $\pi$/3           & $\pi$      & $\pi$       & -2                        &           & band 3        & -$\pi$/3          & $\pi$/3    & $\pi$/3     & 1                         \\ \midrule
\textbf{3.89} & \textbf{$\Gamma$} & \textbf{K} & \textbf{K'} & \textbf{Chern Number (↑)} & \textbf{} & \textbf{1.89} & \textbf{$\Gamma$} & \textbf{K} & \textbf{K'} & \textbf{Chern Number (↑)} \\ \midrule
band 1        & $\pi$/3           & $\pi$      & $\pi$       & 1                         &           & band 1        & $\pi$/3           & $\pi$      & $\pi$       & 1                         \\ \midrule
band 2        & $\pi$             & -$\pi$/3   & -$\pi$/3    & 1                         &           & band 2        & $\pi$             & -$\pi$/3   & -$\pi$/3    & 1                         \\ \midrule
band 3        & $\pi$/3           & $\pi$      & $\pi$       & -2                        &           & band 3        & -$\pi$/3          & $\pi$/3    & $\pi$/3     & 1                         \\ \midrule
\textbf{3.48} & \textbf{$\Gamma$} & \textbf{K} & \textbf{K'} & \textbf{Chern Number (↑)} & \textbf{} & \textbf{1.70} & \textbf{$\Gamma$} & \textbf{K} & \textbf{K'} & \textbf{Chern Number (↑)} \\ \midrule
band 1        & $\pi$/3           & $\pi$      & $\pi$       & 1                         &           & band 1        & $\pi$             & $\pi$      & $\pi$       & 0                         \\ \midrule
band 2        & $\pi$             & -$\pi$/3   & -$\pi$/3    & 1                         &           & band 2        & $\pi$/3           & -$\pi$/3   & -$\pi$/3    & -1                        \\ \midrule
band 3        & $\pi$/3           & $\pi$      & $\pi$       & -2                        &           & band 3        & -$\pi$/3          & $\pi$/3    & $\pi$/3     & 1                         \\ \midrule
\textbf{3.14} & \textbf{$\Gamma$} & \textbf{K} & \textbf{K'} & \textbf{Chern Number (↑)} & \textbf{} & \textbf{1.47} & \textbf{$\Gamma$} & \textbf{K} & \textbf{K'} & \textbf{Chern Number (↑)} \\ \midrule
band 1        & $\pi$/3           & $\pi$      & $\pi$       & 1                         &           & band 1        & $\pi$             & $\pi$      & $\pi$       & 0                         \\ \midrule
band 2        & $\pi$             & -$\pi$/3   & -$\pi$/3    & 1                         &           & band 2        & $\pi$/3           & $\pi$/3    & $\pi$/3     & 0                         \\ \midrule
band 3        & -$\pi$/3          & $\pi$/3    & $\pi$/3     & -2                        &           & band 3        & -$\pi$/3          & -$\pi$/3   & -$\pi$/3    & 0                         \\ \midrule
\textbf{2.87} & \textbf{$\Gamma$} & \textbf{K} & \textbf{K'} & \textbf{Chern Number (↑)} & \textbf{} & \textbf{1.30} & \textbf{$\Gamma$} & \textbf{K} & \textbf{K'} & \textbf{Chern Number (↑)} \\ \midrule
band 1        & $\pi$/3           & $\pi$      & $\pi$       & 1                         &           & band 1        & $\pi$             & $\pi$      & $\pi$       & 0                         \\ \midrule
band 2        & $\pi$             & -$\pi$/3   & -$\pi$/3    & 1                         &           & band 2        & -$\pi$/3          & $\pi$/3    & $\pi$/3     & 1                         \\ \midrule
band 3        & -$\pi$/3          & $\pi$/3    & $\pi$/3     & 1                         &           & band 3        & $\pi$/3           & -$\pi$/3   & -$\pi$/3    & -1                        \\ \midrule
\textbf{2.64} & \textbf{$\Gamma$} & \textbf{K} & \textbf{K'} & \textbf{Chern Number (↑)} & \textbf{} & \textbf{1.08} & \textbf{$\Gamma$} & \textbf{K} & \textbf{K'} & \textbf{Chern Number (↑)} \\ \midrule
band 1        & $\pi$/3           & $\pi$      & $\pi$       & 1                         &           & band 1        & $\pi$             & $\pi$      & $\pi$       & 0                         \\ \midrule
band 2        & $\pi$             & -$\pi$/3   & -$\pi$/3    & 1                         &           & band 2        & -$\pi$/3          & $\pi$/3    & $\pi$/3     & 1                         \\ \midrule
band 3        & -$\pi$/3          & $\pi$/3    & $\pi$/3     & 1                         &           & band 3        & $\pi$/3           & $\pi$      & $\pi$       & 1                         \\ \midrule
\textbf{2.45} & \textbf{$\Gamma$} & \textbf{K} & \textbf{K'} & \textbf{Chern Number (↑)} & \textbf{} & \textbf{0.89} & \textbf{$\Gamma$} & \textbf{K} & \textbf{K'} & \textbf{Chern Number (↑)} \\ \midrule
band 1        & $\pi$/3           & $\pi$      & $\pi$       & 1                         &           & band 1        & -$\pi$/3          & $\pi$      & $\pi$       & -1                        \\ \midrule
band 2        & $\pi$             & -$\pi$/3   & -$\pi$/3    & 1                         &           & band 2        & -$\pi$/3          & $\pi$/3    & $\pi$/3     & 1                         \\ \midrule
band 3        & -$\pi$/3          & $\pi$/3    & $\pi$/3     & 1                         &           & band 3        & $\pi$             & $\pi$      & $\pi$       & 0                         \\ \midrule
\textbf{2.28} & \textbf{$\Gamma$} & \textbf{K} & \textbf{K'} & \textbf{Chern Number (↑)} & \textbf{} &               &                   &            &             &                           \\ \midrule
band 1        & $\pi$/3           & $\pi$      & $\pi$       & 1                         &           &               &                   &            &             &                           \\ \midrule
band 2        & $\pi$             & -$\pi$/3   & -$\pi$/3    & 1                         &           &               &                   &            &             &                           \\ \midrule
band 3        & -$\pi$/3          & $\pi$/3    & $\pi$/3     & 1                         &           &               &                   &            &             &                           \\ \bottomrule
\end{tabular}
\caption{
The $C_{3z}$ symmetry eigenvalues of the high symmetry momentum points, computed at a twist angle of 5.08$^{\circ}$, 4.41$^{\circ}$, 3.89$^{\circ}$, 3.48$^{\circ}$, 3.14$^{\circ}$, 2.87$^{\circ}$, 2.65$^{\circ}$, 2.45$^{\circ}$, 2.28$^{\circ}$, 2.13$^{\circ}$, 2.01$^{\circ}$, 1.89$^{\circ}$, 1.70$^{\circ}$, 1.47$^{\circ}$, 1.30$^{\circ}$, 1.08$^{\circ}$, and 0.89$^{\circ}$.}
\end{table}








\end{document}